\documentclass[12pt ]{article}
\usepackage{jheppub_kim}
\usepackage{amsmath}
\usepackage{amssymb}
\usepackage{dcolumn}
\usepackage{bm}
\usepackage{color}
\usepackage{epsfig}
\usepackage{amsfonts}
\usepackage{graphicx}
\usepackage{subfigure}
\usepackage{multirow}
\newcommand{\be}{\begin{equation}}
\newcommand{\ee}{\end{equation}}
\newcommand{\beq}{\begin{equation}}
\newcommand{\eeq}{\end{equation}}
\newcommand{\bea}{\begin{eqnarray}}
\newcommand{\eea}{\end{eqnarray}}
 
\def\be{\begin{equation}}
\def\ee{\end{equation}}
\def\ba{\begin{eqnarray}}
\def\ea{\end{eqnarray}}
\begin{document}
\title{Holographic Thermodynamics of an Enhanced Charged AdS Black Hole in String Theory's Playground}

\author{Behnam Pourhassan\textsuperscript{1,2},}
\affiliation[1]{Canadian Quantum Research Center 204-3002 32 Ave Vernon, BC V1T 2L7 Canada}
\author{Sareh Eslamzadeh\textsuperscript{1},}
\affiliation[2]{Physics Department, Istanbul Technical University, Istanbul 34469, Turkey }
\author{\.{I}zzet Sakall{\i}\textsuperscript{3}}
\affiliation[3]{Physics Department, Eastern Mediterranean University, Famagusta 99628, North Cyprus via Mersin 10, Turkey  }
\author{and Sudhaker Upadhyay\textsuperscript{4,5}}
\affiliation[4]{Department of Physics, K.L.S. College, Magadh University, Nawada-805110, Bihar,  India}
\affiliation[5]{Visiting Associate, Inter-University Centre for Astronomy and Astrophysics (IUCAA) Pune-411007, Maharashtra, India}
\emailAdd{b.pourhassan@candqrc.ca}
\emailAdd{s.eslamzadeh@stu.umz.ac.ir}
\emailAdd{izzet.sakalli@emu.edu.tr}
\emailAdd{sudhakerupadhyay@gmail.com}
\abstract{
In this paper, we consider an $\alpha^{\prime}$ corrected
Reissner-Nordstr\"{o}m AdS black hole to study thermodynamics. We study the
$P-V$ criticality and thermodynamical stability of the black hole. We obtained
a first-order phase transition, which may be interpreted as the large/small
black hole phase transition. Therefore, we obtained a van der Waals
behaviour and obtained critical points. Finally, we calculate quantum work used to resolve the information loss paradox.
 }

 \maketitle
\section{Introduction}
Strominger and Vafa observed that the microscopic entropy of 5-dimensional  extremal black holes in string theory matches with the macroscopic entropy at
the first order \cite{01}. Beyond the first order,  the macroscopic entropy requires corrections to the superstring  theory effective action, and corresponding black hole solutions are valid to the level of approximation needed. Another way to phrase this is that while most black-hole solutions in string theory adhere only to the effective field theories at the zeroth order, the precision of stringy black hole solutions can be improved by incorporating higher-order terms in $\alpha^{\prime}$ corrections. A modified version of black holes, accounting for corrections associated with string theory, where $\alpha^{\prime}$ represents the string length scale, is an intriguing topic in theoretical physics. The inclusion of higher-order terms in $\alpha^{\prime}$ is crucial as they signify genuine deviations from matter coupled Einstein gravity due to the non-zero string length. The validity of microscopic entropy can only be considered circumstantial without explicit knowledge of complete $\alpha^{\prime}$ corrected black hole solutions and subsequent entropy computation using the action.

The first-order $\alpha^{\prime}$ correction to a family of solutions of the heterotic superstring effective action, which are parallel to solitonic 5-branes with Kaluza-Klein monopoles, is studied \cite{02}. It is demonstrated that this family of $\alpha^{\prime}$ corrected solutions remains invariant under $\alpha^{\prime}$ corrected $T$-duality transformations. Additionally, the first-order $\alpha^{\prime}$ correction to the 4-charge Reissner-Nordstr\"om black hole beyond the near-horizon limit in the heterotic superstring effective action is examined \cite{03}. In this context, the Wald entropy is found to be in agreement with the entropy obtained through microstate counting. Furthermore, the first-order $\alpha^{\prime}$ correction to non-extremal 4-dimensional dyonic Reissner-Nordstr\"om black holes with equal electric and magnetic charges is also discussed in the context of heterotic superstring effective field theory \cite{04}.

In this work, we consider a 4-dimensional $\alpha^{\prime}$ corrected
Reissner-Nordstr\"{o}m AdS black hole. We first study the effect of $
\alpha^{\prime}$ correction, charge parameter and cosmological constant on
the metric function.   To be more precise,  the decrease in charge parameter creates
one coincidence horizon and two separated horizons.  As the charge of the
black hole decreases, the outer horizons of the Reissner-Nordstr\"{o}m AdS
black hole with/without $\alpha^{\prime}$ correction have more overlap.
With the decrease in cosmological constant, the outer horizons disappear but the inner horizon, which is
constructed due to the existence of $\alpha^{\prime}$ correction, exists
for any cosmological constant.  The increasing $\alpha^{\prime}$  widens the separation of two outer horizons and the inner horizon radius
bigger. We also study the thermodynamics of the black hole.
Incidentally, the existence of $\alpha^{\prime}$ prevents the outer horizon's final temperature from zero, but it doesn't affect the critical mass of the black hole. With increasing $\alpha^{\prime}$ correction, the unstable phase is gradually corrected, and the heat capacity becomes continuously positive. However,  $\alpha^{\prime}$  correction causes instability in the final evaporation stage when the black hole mass approaches the critical mass. Finally, we compute quantum work for this black hole.
The quantum work is
represented by a unitary information preserving process, which is the dual picture of an $\alpha^{\prime}$ corrected Reissner-Nordstr\"{o}m AdS black hole. We find that an $\alpha^{\prime}$ corrected Reissner-Nordstr\"{o}m AdS black hole is unstable through a first-order phase transition and will evaporate at
the final state.

The paper is organized in the following manner.
In section \ref{sec1}, we focus on $\alpha^{\prime}$ corrected  4-
dimensional Reissner-Nordstr\"{o}m AdS black hole
solution and discuss the horizon structure. In section
\ref{sec2}, we study the thermodynamics of black holes
by calculating Hawking temperature and entropy. Section \ref{sec4} is
devoted to calculating quantum work for the $\alpha^{\prime}$ corrected
Reissner-Nordstr\"{o}m AdS black hole. Finally, we conclude the results in
the last section.

\section{Black hole solution}\label{sec1}
\begin{figure}[ht] \label{fig4}
		\centering
		\includegraphics[height=3.5 cm,width=5 cm]  {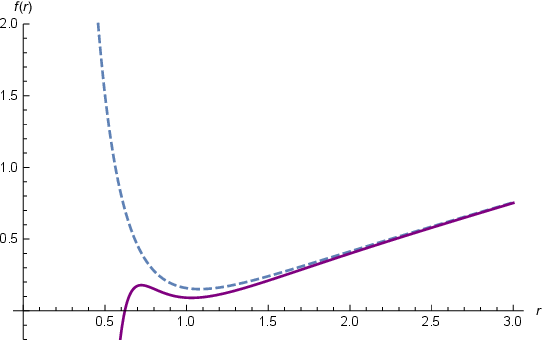}
		
		\vspace*{0.3cm}
		\includegraphics[height=3.5 cm,width=5 cm]  {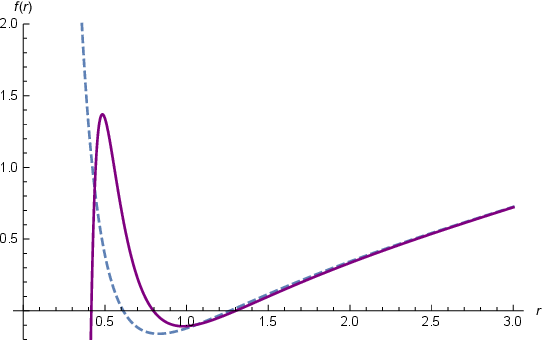}
		\hspace*{0.1cm}
		\includegraphics[height=3.5 cm,width=5 cm]  {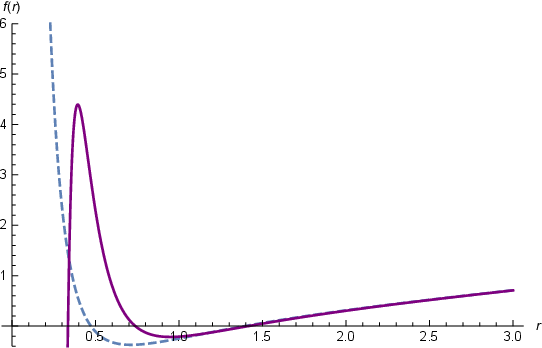}
		\vspace*{0.5cm}
		\caption{{ $f(r)$ versus of $r$ with different $p$ for the $\alpha^{\prime}$ corrected Reissner-Nordstr\"{o}m AdS black hole. Plots have been depicted with $p=1.5,1.4,1.3,1.2$ from left to right. Dashed line shows the Reissner-Nordstr\"{o}m black hole without correction ($\alpha^{\prime}=0$). As $p$ decreases, two separated horizons are created. There exists critical $p$ for which we have one coincidence horizon. Plots have been depicted with $M=1, \alpha^{\prime}=M^2 (a=1), \Lambda=-0.1$.}}
		\end{figure}

\begin{figure}[ht]
		\centering
		\includegraphics[height=3.5 cm,width=5 cm]  {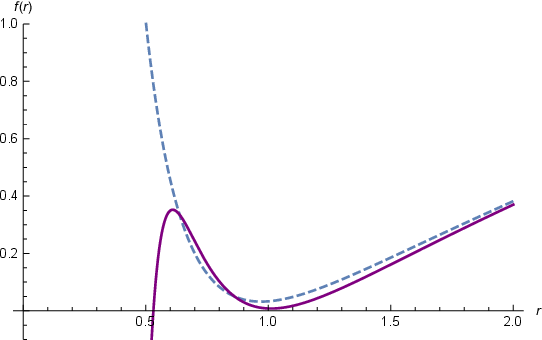}
		\hspace*{0.1cm}
		\includegraphics[height=3.5 cm,width=5 cm] {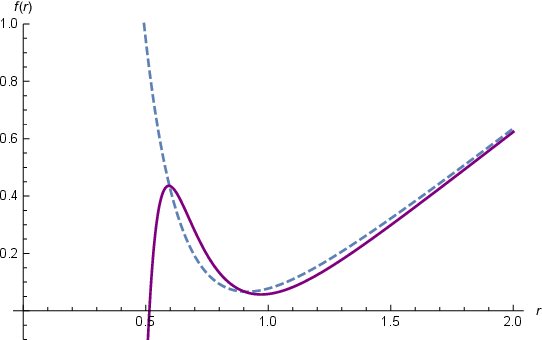}\\
		\vspace*{0.3cm}
		\includegraphics[height=3.5 cm,width=5cm]  {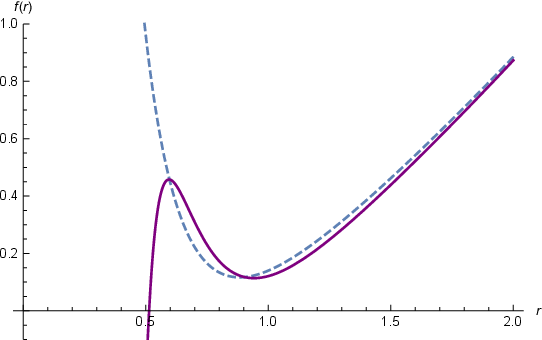}
		\hspace*{0.1cm}
		\includegraphics[height=3.5 cm,width=5cm]  {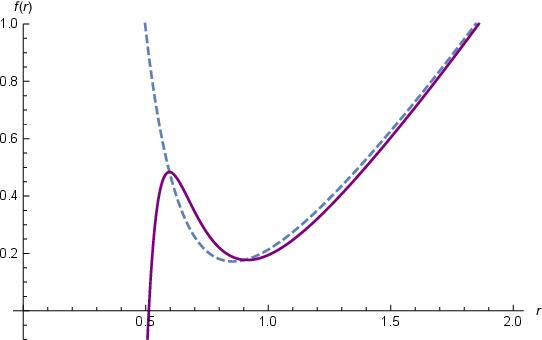}
		\vspace*{0.5cm}
		\caption{{ $f(r)$ versus of $r$ with different $\Lambda$ for the $\alpha^{\prime}$ corrected Reissner-Nordstr\"{o}m AdS black hole. Plots have been depicted with $\Lambda=-0.1,-0.3,-0.5,-0.7$ from left to right. Dashed line shows the Reissner-Nordstr\"{o}m black hole without correction ($\alpha^{\prime}=0$). As $\Lambda$ decreases, outer horizons disappear. Plots have been depicted with $M=1, \alpha^{\prime}= M^2 (a=1), p\simeq1.4$.}}
		\label{fig5}
		\end{figure}

\begin{figure}[ht]
 		\centering
		\includegraphics[height=3.5 cm,width=5cm]{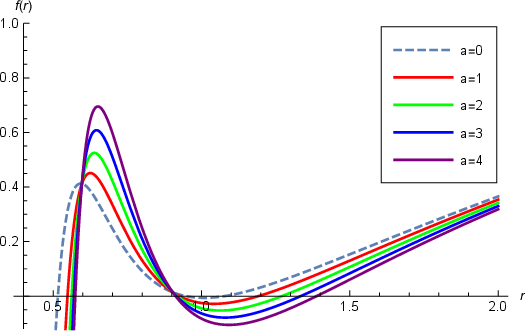}
		\caption{$f(r)$ versus of $r$ with different $\alpha^{\prime}=a M^2$ for the $\alpha^{\prime}$ corrected Reissner-Nordstr\"{o}m AdS black hole.}
		\label{fig6}
		\end{figure}

A 4-dimensional Reissner-Nordstr\"{o}m AdS black hole of mass $M$ with the
$\alpha^{\prime}$ correction given by the following metric \cite{1},
\begin{equation}\label{metric001}
ds^{2}=N^{2}f(r)dt^{2}-\frac{dr^{2}}{f(r)}-r^{2}(d\theta^{2}+\sin^{2}\theta
d\varphi^{2}),
\end{equation}
with
\begin{equation}
N^{2}=1+\alpha^{\prime}\frac{p^{2}}{8r^{4}},
\end{equation}
\begin{equation}\label{frAds}
f(r)=1-\frac{2M}{r}+\frac{r^{2}}{l^{2}}+\frac{p^{2}}{2r^{2}}-
\alpha^{\prime}\frac{p^{2}}{4r^{4}}\left[1-\frac{3M}{2r}+\frac{11p^{2}}
{40r^{2}}\right],
\end{equation}
where $p$ is a physical constant related to the black hole charges, and
$l$, at $4D$ space-time, is the AdS radius, which is related to the cosmological constant via
\begin{equation}
\Lambda=-\frac{3}{l^{2}}.
\end{equation}
We look into the structure of the $\alpha^{\prime}$ corrected Reissner-Nordstr\"{o}m AdS black hole, and we study the effect of this correction on the metric. According to equation (\ref{frAds}), we check the metric dependency on $p$, $\alpha^{\prime}$ and, $\Lambda$. In Figure \ref{fig4}, we plot $f(r)$ versus radius with different $p$. Here, we find that the decrease in $p$   creates one coincidence horizon and two separated horizons. Hence, comparing the $\alpha^{\prime}$ corrected Reissner-Nordstr\"{o}m AdS black hole with Reissner-Nordstr\"{o}m AdS black hole shows the existence of $\alpha^{\prime}$ correction causes for any content of the charge, ever less than $p_{cri}=\sqrt{2}M$, there will be at least one horizon. Moreover, as the charge of the black hole decreases, the outer horizons of the Reissner-Nordstr\"{o}m AdS black hole with/without $\alpha^{\prime}$ correction have more overlap. In Figure \ref{fig5}, we fix the black hole's charge to the critical charge.  As $\Lambda$ decreases, outer horizons disappear, but the inner horizon, which is constructed due to the existence of $\alpha^{\prime}$ correction, exists for any cosmological constant. In the following, we fix the mass and charge of the black hole, and we probe the effect of the $\alpha^{\prime}$ parameter. As shown in Figure \ref{fig6}, increasing $\alpha$ (increasing $a$) makes two outer horizons more separated and the inner horizon's radius bigger.

The  area of the event horizon  for the black hole  in terms of horizon radius $r_h$
is given by
\begin{equation}\label{Area}
A_{h}=4\pi r_{h}^{2},
\end{equation}
and the black hole volume is
\begin{equation}\label{V}
V=\frac{4}{3}\pi r_{h}^{3}.
\end{equation}

\section{Thermodynamics}\label{sec2}

\begin{figure}[!ht]
		\centering
		\includegraphics[height=4.5 cm,width=6.5 cm]  {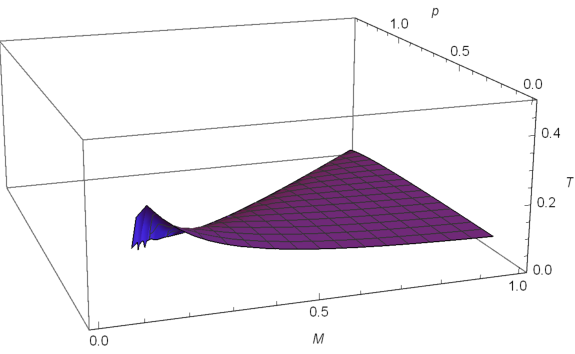}
		\quad
		\includegraphics[height=4.5 cm,width=6.5 cm] {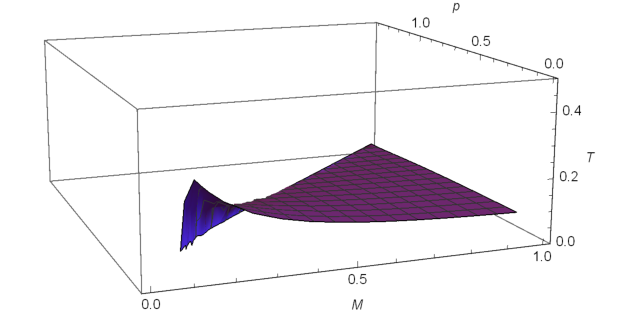}
		\caption{{The behaviour of temperature versus the mass and the charge of the $\alpha^{\prime}$ corrected Reissner-Nordstr\"{o}m AdS black hole. Left plot has been depicted with $\alpha^{\prime}=2 M^2$ and right one with $\alpha^{\prime}=5 M^2$ while $\Lambda=-0.1$. Increasing  $\alpha^{\prime}$ predicts less temperature, also, less mass and charge for the remnant of the black hole.}}
		\label{Fig7}
	\end{figure}

\begin{figure}[!ht]
		\centering
		\includegraphics[height=4.5 cm,width=6.5 cm]  {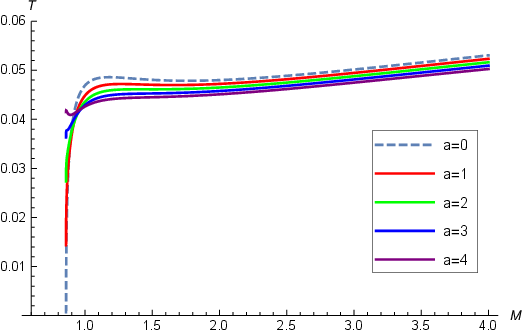}
		\quad
		\includegraphics[height=4.5 cm,width=6.5 cm] {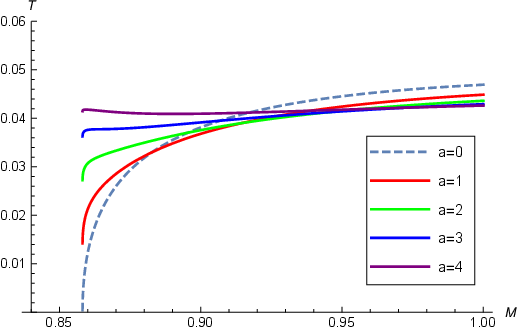}
		\caption{{The behavior of temperature versus the mass for the $\alpha^{\prime}$ corrected Reissner-Nordstr\"{o}m AdS black hole. Plots have been depicted with $a=0,1,2,3,4$ ($\alpha^{\prime}=a M^2$) and $\Lambda=-0.1$ while $p<p_{cri}$.}}
		\label{Fig8}
		\end{figure}

We calculate the Hawking temperature of this kind of black hole based on the famous formula between the surface gravity of the event horizon and the temperature that has been given by \cite{2}
	\begin{equation}\label{Tem}
	T=\frac{1}{4\pi}\Big{[}\frac{1}{\sqrt{-g_{tt}g_{rr}}}\frac{dg_{tt}}{dr}\Big{]}_{r_{h}}.
	\end{equation}
Considering the metric \ref{frAds}, we find the horizons of the $\alpha^{\prime}$ corrected Reissner-Nordstr\"{o}m AdS black hole, and we calculate the temperature of the outer horizon, $r_{h}$, as follows,

\begin{equation}\label{TemAds}
	T=\frac{l^2 \left(-10 p^2 r_{h} \left(15 \alpha^{\prime}  M+8 r_{h}^3-8 \alpha^{\prime} r_{h}\right)+160 M r_{h}^5+33 \alpha^{\prime}  p^4\right)+160 r_{h}^8}{80 \sqrt{2} \pi  l^2 r_{h}^7 \sqrt{\frac{\alpha^{\prime}  p^2}{r_{h}^4}+8}},
	\end{equation}
where $r_{h}$ is the largest root of $f(r)=0$ using the equation (\ref{frAds}).
Also, we illustrate it in Figures \ref{Fig7} and \ref{Fig8}. One can see in Figure \ref{Fig7} that decreasing the mass and charge of the black hole, the temperature of the  $\alpha^{\prime}$ corrected Reissner-Nordstr\"{o}m AdS black hole increases until it reaches a maximum and after that decreases to a finite temperature with finite mass and charge. To research about the effect of $\alpha^{\prime}$, we plot Figure \ref{Fig8} with different $a$. The most important result is that the existence of $\alpha^{\prime}$ prevents the outer horizon's final temperature from zero, but it doesn't affect the critical mass of the black hole.

\begin{figure}[ht]
		\centering
		\includegraphics[height=4.5 cm,width=6.5 cm]  {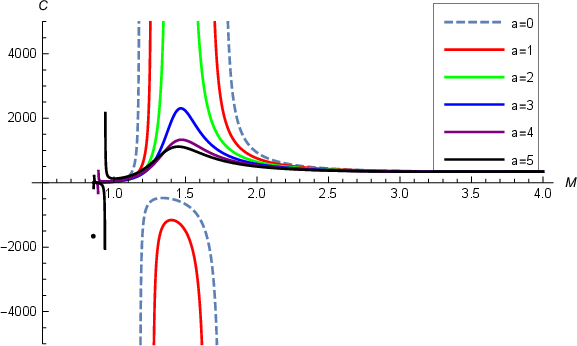}
		
		\caption{{The behavior of temperature versus the mass for the $\alpha^{\prime}$ corrected Reissner-Nordstr\"{o}m AdS black hole. Plots have been depicted with $a=0,1,2,3,4$ ($\alpha^{\prime}=a M^2$) and $\Lambda=-0.1$ while $p<p_{cri}$.}}
		\label{Fig9}
		\end{figure}

We consider the entropy of the $\alpha^{\prime}$ corrected Reissner-Nordstr\"{o}m AdS black hole as a
	\begin{equation}\label{Ent}
	S=\pi r_{h}^{2},
	\end{equation}
which $r_{h}$ is the largest root of the equation (\ref{frAds}) as a outer horizon. To investigate the heat capacity,
\begin{equation}\label{C}
C=T\frac{dS}{dT},
\end{equation}
we may approximate in two regions as,
\begin{eqnarray}\label{C1}
C=\begin{array}{cc}
    T\frac{dS}{dM}\left(\frac{dT}{dM}\right)^{-1}, & \left(\frac{dT}{dM}\gg\frac{dT}{dq}\right)\\
    T\frac{dS}{dq}\left(\frac{dT}{dq}\right)^{-1}, & \left(\frac{dT}{dM}\ll\frac{dT}{dq}\right).
  \end{array}
\end{eqnarray}
So, we employ the first condition of equation (\ref{C1}) and present the results of our calculations in Figure \ref{Fig9}. As expected for the Reissner-Nordstr"{o}m AdS black hole, there exists an unstable thermodynamic phase ($C<0$) when $r_{h}$ falls within a specific range. The black hole is in a stable thermodynamic phase ($C>0$) at the final stage of evaporation \cite{Al14} (See dashed blue curve in Figure \ref{Fig9}). A notable outcome is that when we introduce $\alpha^{\prime}$ corrections and increase their values, the thermodynamically unstable phase gradually corrects itself, and the heat capacity becomes continuously positive. Conversely, as depicted in the right panel of Figure \ref{Fig9}, this type of correction induces instability in the final evaporation stage as the black hole mass approaches the critical mass. Additionally, the larger the value of $\alpha^{\prime}$, the more phase transitions are predicted.

\begin{figure}[!ht]
 		\centering
		\includegraphics[width=75 mm]{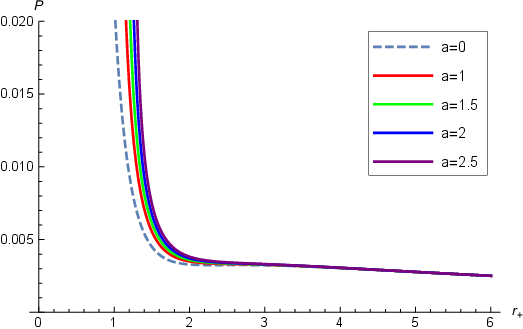}
		\caption{$P-V$ diagram of the $\alpha^{\prime}$ corrected Reissner-Nordstr\"{o}m AdS black hole for various $\alpha^{\prime}$ ($\alpha^{\prime}=a M^2$) per critical temperature. }
		\label{Fig10}
		\end{figure}

\begin{figure}[!ht]
 		\centering
		\includegraphics[width=75 mm]{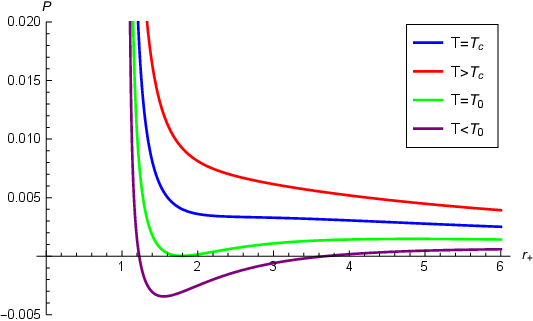}
		\caption{$P-V$ diagram of the $\alpha^{\prime}$ corrected Reissner-Nordstr\"{o}m AdS black hole per various temperature. Plots have been depicted with $\alpha^{\prime}=1.5 M^2$.}
		\label{Fig11}
		\end{figure}

Eventually, we will investigate the effect of the $\alpha^{\prime}$ correction on the Van der Waals equation. According to the outer horizon of the metric (\ref{frAds}), we derive the mass of the black hole as follows
	\begin{equation}\label{Mass}
	M=\frac{l^2 \left(-11 \alpha^{\prime}  p^4+40 p^2 \left(2 r_{h}^4-\alpha^{\prime}  r_{h}^2\right)+160 r_{h}^6\right)+160 r_{h}^8}{20 l^2 \left(16 r_{h}^5-3 \alpha^{\prime}  p^2 r_{h}\right)}.
	\end{equation}
Hence, the black hole pressure and its conjugate variable, the black hole volume, are given by
	\begin{equation}\label{Pre2}
	P=\frac{3}{8\pi l^{2}},\hspace{1cm}
	V=\frac{4}{3}\pi r_{h}^{3}.
	\end{equation}
In this case, the first law of thermodynamics is modified as,
	\begin{equation}\label{First-law2}
dM=TdS+\Phi dq+VdP+Ad\alpha^{\prime},
	\end{equation}
where $\Phi$ is the corrected electrostatic potential conjugate to the black hole charge, while $A$ is a thermodynamics variable conjugate to the correction parameter.
Therefore, the thermodynamics relations read as,
\begin{align}\label{T-R}
&\left(\frac{\partial S}{\partial M}\right)_{q,\alpha^{\prime}}=\frac{1}{T},&&
\left(\frac{\partial S}{\partial q}\right)_{M,\alpha^{\prime}}=-\frac{\Phi}{T},&&
\left(\frac{\partial S}{\partial \alpha^{\prime}}\right)_{M,q}=-\frac{A}{T}.
\end{align}
Based on equation (\ref{Pre2}), if we suppose the cosmological constant has not been allowed to vary, $P$ is treated as a constant. Moreover, for a fixed $\alpha^{\prime}$ and a fixed charge, we derive the equation of state for the $\alpha^{\prime}$ corrected Reissner-Nordstr\"{o}m AdS black hole concerning equations (\ref{TemAds}), (\ref{Mass}), and (\ref{Pre2}) as follows
	\begin{equation}\label{State}
	P=\frac{\frac{10 \pi T}{v}\sqrt{1+\frac{2p^2 \alpha^{\prime}}{v^4}}\Big{(}v^4-3 p^2 \alpha^{\prime} \Big{)}-5v^2+10p^2+\frac{66p^2 \alpha^{\prime2}}{v^8}-\frac{60p^4 \alpha^{\prime2}}{v^6}-\frac{20p^4 \alpha^{\prime}}{v^4}}{10 \pi (v^4-7 p^2 \alpha^{\prime})},
	\end{equation}
where, $v$, is the specific volume with $v=2l_{Pl}^2 r_{h}$. Besides, one can find the critical point in the condition
	\begin{equation}\label{Cond}
	\frac{\partial{P}}{\partial{v}}=0,\hspace{1cm}
	\frac{\partial^2{P}}{\partial{v^2}}=0.
	\end{equation}
We present the $P-V$ plot in Figure \ref{Fig10} for the $\alpha^{\prime}$-corrected Reissner-Nordstr"{o}m AdS black hole per critical temperature. As demonstrated, increasing $\alpha^{\prime}$ reduces the oscillatory component of the curves. In other words, increasing $\alpha^{\prime}$ leads to phase transitions occurring over smaller changes in the volume of the black hole. Additionally, we maintain $\alpha^{\prime}$ constant and illustrate decreasing temperature from top to bottom in Figure \ref{Fig11}. If the temperature exceeds the critical temperature $T_{c}$, no phase transition occurs. When the temperature falls between $T_{c}$ and $T_{0}$, the behavior of the black hole resembles that of a Van der Waals gas, exhibiting an inflection point. Finally, if the temperature is below $T_{0}$, the pressure becomes negative for certain values of $r_{+}$. As expected, in the absence of $\alpha^{\prime}$ correction, we recover the equation of state for a charged AdS black hole, as can be found in various references, such as Ref. \cite{Al14, Kub12},
	\begin{equation}\label{StateAds}
	P=\frac{T}{v}-\frac{1}{2\pi v^2}+\frac{p^2}{\pi v^4}.
	\end{equation}
	
\section{Quantum work}\label{sec4}
Black hole entropy, at quantum scales, is related to the microstates. Hence, a change in the black hole microstates yields a change in the entropy. So, one
can express the change in the quantum corrected entropy of the Reissner-Nordstr\"{o}m AdS
black hole as
\begin{equation}\label{Delta S}
\Delta S=S(r_{h2}) - S(r_{h1}).
\end{equation}
This change in the entropy and the first law of thermodynamics \cite{Do11} can be used to obtain a change in the internal energy of the Reissner-Nordstr\"{o}m AdS
black hole.\\
The internal energy of $\alpha^{\prime}$ corrected Reissner-Nordstr\"{o}m AdS black hole is calculated as following,
\begin{equation}\label{In.E}
E=\frac{X_E}{120 l^2 r_{h}^5 \left(r_{h}^4-112 \alpha^{\prime}  p^2\right) \left(16 r_{h}^4-3 \alpha^{\prime}  p^2\right) \sqrt{\frac{\alpha^{\prime}  p^2}{r_{h}^4}+8}},
\end{equation}
where
\begin{eqnarray}\label{Energy}
X_E&=& -480\alpha^{\prime}  p^2 r_{h}^{12} \left(224 \sqrt{\frac{\alpha^{\prime}  p^2}{r_{h}^4}+8}-775 \sqrt{\frac{64 \alpha^{\prime}  p^2}{r_{h}^4}+2}\right)\nonumber\\
&-&480r_{h}^{16} \left(16 \sqrt{\frac{64 \alpha ^{\prime} p^2}{r_{h}^4}+2}-2 \sqrt{\frac{\alpha^{\prime}  p^2}{r_{h}^4}+8}\right)\nonumber\\
&-&161280 \alpha^{\prime2} r_{h}^8 p^4 \sqrt{\frac{64 \alpha^{\prime}  p^2}{r_{h}^4}+2}\nonumber\\
&+&320 r_{h}^{14}l^2 \left(19 \sqrt{\frac{\alpha^{\prime}  p^2}{r_{h}^4}+8}-8 \sqrt{\frac{64 \alpha^{\prime}  p^2}{r_{h}^4}+2}\right)\nonumber\\
&+&1584 p^8 l^2 \alpha^{\prime3} \left(8192 \sqrt{\frac{\alpha^{\prime}  p^2}{r_{h}^4}+8}-\sqrt{\frac{64 \alpha^{\prime}  p^2}{r_{h}^4}+2}\right)\nonumber\\
&+&20 \alpha^{\prime2}  p^4 r_{h}^6 \left(790080  \sqrt{\frac{\alpha^{\prime}  p^2}{r_{h}^4}+8}-1158  \sqrt{\frac{64 \alpha^{\prime}  p^2}{r_{h}^4}+2}\right)\nonumber\\
&+&2 \alpha^{\prime}  p^4 r_{h}^8 \left(632287 \sqrt{\frac{\alpha^{\prime}  p^2}{r_{h}^4}+8}-30800 \sqrt{\frac{64 \alpha^{\prime}  p^2}{r_{h}^4}+2}\right)\nonumber\\
&+&80 p^2 \alpha^{\prime}r_{h}^{10} \left(-4431   \sqrt{\frac{\alpha^{\prime}  p^2}{r_{h}^4}+8}+1542  \sqrt{\frac{64 \alpha^{\prime}  p^2}{r_{h}^4}+2}\right)\nonumber\\
&+&80 p^2 r_{h}^{12}\left(16 \sqrt{\frac{64 \alpha^{\prime}  p^2}{r_{h}^4}+2}-506 \sqrt{\frac{\alpha^{\prime}  p^2}{r_{h}^4}+8}\right)\nonumber\\
&+&\alpha^{\prime3} p^6 r_{h}^2 \left(-983040   \sqrt{\frac{\alpha^{\prime}  p^2}{r_{h}^4}+8}+1920   \sqrt{\frac{64 \alpha^{\prime}  p^2}{r_{h}^4}+2}\right)\nonumber\\
&+&\alpha^{\prime2} p^6 r_{h}^4  \left(2571 \sqrt{\frac{64 \alpha^{\prime}  p^2}{r_{h}^4}+2}-23148128 \sqrt{\frac{\alpha^{\prime}  p^2}{r_{h}^4}+8}\right).
\end{eqnarray}
Hence, we can obtain the change in internal energy as
\begin{equation}\label{Delta E}
\Delta E=E(r_{h2}) - E(r_{h1}).
\end{equation}
In the final stage, to study
the amount of quantum work, we need to derive Helmholtz free energy of the black hole, we have to consider the modified Smarr-Gibbs-Duhem formula for a $\alpha^{\prime}$ corrected Reissner-Nordstr\"{o}m AdS black hole as follows
\begin{equation}\label{Sm}
M=2TS-2PV+\Phi Q+A\alpha^{\prime}.
\end{equation}
Comparing the first law of thermodynamics, equation (\ref{First-law2}), with Smarr formula, equation (\ref{Sm}), we identify the Helmholtz free energy of the black hole as mentioned in Ref. \cite{Maj17} as a
	\begin{equation}\label{H.E}
	F=E-TS.
	\end{equation}	
As a result, we find Helmholtz free energy of the $\alpha^{\prime}$ corrected Reissner-Nordstr\"{o}m AdS black hole as follows
\begin{equation}\label{H.E.F}
F=\frac{X_{F}}{960 l^2 r_{h}^5 \left(r_{h}^4-112 \alpha^{\prime}  p^2\right) \left(16 r_{h}^4-3 \alpha^{\prime}  p^2\right) X_{1}},
	\end{equation}
where
\begin{align}\label{Hel.Energy}
X_F=&-480 \Bigl(r_{h}^{16} \Bigl(-16 X_{1}+128 X_{2}+48 \sqrt{2}\Bigr)+\alpha^{\prime}  p^2 r_{h}^{12} \Bigl(1792 X_{1}-6200 X_{2}-5397 \sqrt{2}\Bigr)\nonumber\\
&+336 \alpha^{\prime2} p^4 r_{h}^8 \Bigl(8 X_{2}+7\sqrt{2}\Bigr)+l^2\Bigl(-2560 r_{h}^{14} \Bigl(-19 X_{1}+8 X_{2}+3 \sqrt{2}\Bigr)\nonumber\\
&-1584 \alpha^{\prime3} p^8 \Bigl(-65536 X_{1}+8 X_{2}+7 \sqrt{2}\Bigr)\Bigr)\nonumber\\
&+160 p^2 l^2 r_{h}^{10} \Bigl(5385 \sqrt{2} \alpha^{\prime} -17724 \alpha^{\prime}  X_{1}+6168 \alpha^{\prime}  X_{2}+r_{h}^2 \Bigl(-2024 X_{1}+64 X_{2}+24 \sqrt{2}\Bigr)\Bigr)\nonumber\\
&-8 l^2 \alpha^{\prime}  p^4 r_{h}^6 \Bigl(20205 \sqrt{2} \alpha^{\prime} -15801600 \alpha^{\prime}  X_{1}+23160 \alpha^{\prime}  X_{2}\nonumber\\
&+r_{h}^2 \Bigl(-1264574 X_{1}+61600 X_{2}+53820 \sqrt{2}\Bigr)\Bigr)\nonumber\\
&+3 l^2 \alpha^{\prime2} p^6 r_{h}^2 \Bigl(1920 \Bigl(7 \sqrt{2} \alpha^{\prime} -4096 \alpha^{\prime}  X_{1}+8 \alpha^{\prime}  X_{2}\Bigr)\nonumber\\
&+r_{h}^2 \Bigl(-185185024 X_{1}+20568 X_{2}+17953 \sqrt{2}\Bigr)\Bigr)\Bigr),
\end{align}
with
\begin{eqnarray}
X_{1}&\equiv&\sqrt{\frac{\alpha^{\prime}  p^2}{r_{h}^4}+8},\nonumber\\
X_{2}&\equiv&\sqrt{\frac{64 \alpha^{\prime}  p^2}{r_{h}^4}+2}.
\end{eqnarray}
Therefore, the change of Helmholtz free energy can be expressed as,
\begin{equation}\label{Delta F}
\Delta F=F(r_{h2}) - F(r_{h1}).
\end{equation}
Now, the quantum work ($W_{Q}$) is expressed as \cite{JHEP},
\begin{equation}\label{work}
W_{Q}=e^{\frac{\Delta F}{T}},
\end{equation}
where $T$ is given by the equation (\ref{TemAds}).
In this scenario, a dual depiction of black hole geometry is articulated using superconformal field theory, offering a potential resolution to the information loss paradox. The black hole geometry of an $\alpha^{\prime}$ corrected Reissner-Nordstr"{o}m AdS black hole emits Hawking radiation, leading to the information loss paradox in quantum thermodynamics. This paradox may find resolution through the quantum behavior presented in the dual picture of a superconformal field theory. Consequently, an $\alpha^{\prime}$ corrected Reissner-Nordstr"{o}m AdS black hole is described by a unitary superconformal field theory. This framework enables the study of black hole geometry through quantum non-equilibrium thermodynamics, achievable by computing the quantum work of an $\alpha^{\prime}$ corrected Reissner-Nordstr"{o}m AdS black hole. This quantum work is symbolized by a unitary information-preserving process, serving as the dual depiction of the black hole. The non-equilibrium quantum thermodynamics on the alternate side facilitates the loss of mass from an $\alpha^{\prime}$ corrected Reissner-Nordstr"{o}m AdS black hole through a unitary process. Demonstrating that an $\alpha^{\prime}$ corrected Reissner-Nordstr"{o}m AdS black hole is unstable via a first-order phase transition reveals its eventual evaporation. Consequently, information leakage from an $\alpha^{\prime}$ corrected Reissner-Nordstr"{o}m AdS black hole during its final stages of evaporation is conceivable through the unitary information-preserving process. Thus, employing non-equilibrium quantum thermodynamics offers a pathway to resolving the information loss paradox within the given system.

\section{Conclusion}\label{Con}
In this work, we have considered a 4-dimensional $\alpha^{\prime}$ corrected  Reissner-Nordstr\"{o}m AdS black hole solution. Furthermore, we studied  the effects of
$\alpha^{\prime}$ correction, charge parameter and cosmological constant on the solution.   Here, we have found that the decrease in the charge parameter creates
one coincidence horizon and two separated horizons.  As the charge of the
black hole decreases, the outer horizons of the Reissner-Nordstr\"{o}m AdS
black hole with/without $\alpha^{\prime}$ correction have more overlap.
When the cosmological constant decreases, the outer horizons disappear but the inner horizon, which is
constructed due to the existence of $\alpha^{\prime}$ correction, exists
for any cosmological constant. Interestingly, the increasing $\alpha^{\prime}$  widens the separation of two outer horizons, and the inner horizon radius
increases.

Additionally, we have discussed the thermodynamics of the black hole.
Here, we have found that the existence of $\alpha^{\prime}$ prevents the outer horizon's final temperature from zero, but it doesn't affect the critical mass of the black hole. With increasing $\alpha^{\prime}$ correction, the unstable phase is gradually corrected, and the heat capacity becomes continuously positive. On the other hand,  $\alpha^{\prime}$  correction causes instability in the final evaporation stage when the black hole mass approaches the critical mass. Finally, we have derived quantum work for this black hole.
The quantum work is
represented by a unitary information preserving process, which is the dual picture of an $\alpha^{\prime}$ corrected Reissner-Nordstr\"{o}m AdS black hole. We have observed that an $\alpha^{\prime}$ corrected Reissner-Nordstr\"{o}m AdS black hole is unstable through a first-order phase transition and will evaporate at
the final state.

\section*{Authors' Contributions}
All authors have the same contribution.
\section*{Data Availability}
No data are available.
\section*{Conflicts of Interest}
The authors declare that there is no conflict of interest.
\section*{Ethical Considerations}
The authors have diligently addressed ethical concerns, such as informed consent, plagiarism, data fabrication, misconduct, falsification, double publication, redundancy, submission, and other related matters.
\section*{Funding}
This research did not receive any grant from funding agencies in the public, commercial, or non-profit sectors.

\end{document}